\documentclass[aps,prl,reprint]{revtex4-1}
\usepackage[english]{babel}
\usepackage[utf8]{inputenc}
\usepackage{xcolor}
\usepackage{physics}
\usepackage[colorinlistoftodos, color=green!40, prependcaption]{todonotes}
\usepackage{amsthm,amsmath,amssymb}
\usepackage{mathtools}
\usepackage{physics}
\usepackage{graphicx}
\usepackage[left=15mm,right=15mm,top=20mm,columnsep=15pt]{geometry} 
\usepackage{adjustbox}
\usepackage{placeins}
\usepackage[T1]{fontenc}
\usepackage{lipsum}
\usepackage{csquotes}
\usepackage{float}
\usepackage[pdftex, pdftitle={Article}, pdfauthor={Author}]{hyperref} 

\begin{document}

\title{Capillary surfers: wave-driven particles at a {\textcolor{black}{vibrating}} fluid interface}

\author{Ian Ho$^{1,\dagger}$, Giuseppe Pucci$^{1,2,\dagger}$, Anand U. Oza$^{3}$, Daniel M. Harris$^{1}$}

  \email[Correspondence email address: ]{daniel_harris3@brown.edu.\\ $^\dagger$Co-first author.}

  \affiliation{$^{1}$School of Engineering, Brown University, 184 Hope Street, Providence, Rhode Island 02912, USA}
  
  \affiliation{$^{2}$Univ Rennes, CNRS, IPR (Institut de Physique de Rennes)—UMR 6251, F-35000 Rennes, France}
  
  \affiliation{$^{3}$Department of Mathematical Sciences \& Center for Applied Mathematics and Statistics, New Jersey Institute of Technology, Newark, New Jersey 07102, USA}

\date{\today} 

\begin{abstract}

We present an experimental study of capillary surfers, a new fluid-mediated active system that bridges the gap between dissipation- and inertia-dominated regimes. Surfers are wave-driven particles that self-propel and interact on a fluid interface via an extended field of surface waves. A surfer's speed and interaction with its environment can be tuned broadly through the particle, fluid, and vibration parameters. The wave nature of interactions among surfers allows for multistability of interaction modes and promises a number of novel collective behaviors.
\end{abstract}

\maketitle

Active matter systems have recently attracted considerable interest for the possibility of extending statistical mechanics to incorporate non-equilibrium phenomena, as their constituents locally consume energy in order to move or exert forces on each other~\cite{ramaswamy2010mechanics,marchetti2013hydrodynamics}. There has been extensive work on dissipation-~\cite{saintillan2007orientational,Sokolov_viscosity,sanchez2012spontaneous,Goldstein_Turbulence,Palacci_Light,golestanian2012collective,Solon2015,Brosseau2019,huber2018emergence,doostmohammadi2018nematic,Soni2019,Bricard_Rollers,Driscoll_Rollers,Yeo_Rollers} 
and inertia-dominated~\cite{lissaman1970formation,weihs1973hydromechanics,AttanasiNatPhys,becker2015hydrodynamic,oza2019lattices,Scholz2018,Mandal2019} 
active systems (e.g. bacterial suspensions and fish schools, respectively), 
but our understanding of the intermediate regime~\cite{Banerjee2017} between these two extremes is currently limited~\cite{klotsa2019above}. Self-propelled particles are natural or artificial particles that generally represent the constituents of such active systems, as they convert energy from the environment into directed motion~\cite{bechinger2016active}.

Vibrating platforms are suitable sources of diffuse energy to observe macroscopic self-propelled particles~\cite{dorbolo2005dynamics,couder2005walking} and their collective behavior~\cite{narayan2007long,aranson2007swirling,kudrolli2008swarming,deseigne2010collective,Workamp_Disks,saenz2018spin}. 
Indeed, solid asymmetric particles can self-propel with an internal source of energy~\cite{giomi2013swarming,deblais2018boundaries} or 
by rectifying the mechanical vibrations of the platform on which they move~\cite{yamada2003coherent,dorbolo2005dynamics,koumakis2016mechanism}.  In particular, such  ``dry'' active systems~\cite{chate2019dry} represent important and successful platforms to demonstrate collective behavior  
in table-top experiments and to evaluate theoretical predictions.  On the other hand, fluid interfaces are convenient platforms for the self-propulsion of natural~\cite{bush2006walking} and artificial bodies, including solid particles~\cite{nagayama2004theoretical,snezhko2009self,chung2009electrowetting,grosjean2015remote,yang2019passive}, drops~\cite{couder2005walking,pucci2011mutual,pucci2015faraday,ebata2015swimming} and small-scale robots~\cite{chen2018controllable}. 
When the fluid interface is vibrating, millimetric droplets may self-propel by bouncing on the sloped surface wave they generate in select parameter regimes~\cite{couder2005walking,bush2020pilot}, and floating drops may be deformed and driven by surface waves~\cite{pucci2011mutual,pucci2015faraday,ebata2015swimming}. For floating drops, propulsion mechanisms based on asymmetric vortex generation~\cite{ebata2015swimming} and wave radiation pressure~\cite{pucci2015faraday} have been proposed, but the drops are difficult to manipulate directly due to their self-adaptive nature. The study of \textcolor{black}{capillary} wave-driven bodies at the fluid interface is relevant to the propulsion of~\cite{bush2006walking,steinmann2018unsteady,roh2019honeybees} and interactions between~\cite{wilcox1972communication,bleckmann1994stimulus} water-walking insects, controlled particle transport and patterning at fluid interfaces \cite{wright2003patterning,falkovich2005floater,punzmann2014generation}, and novel propulsion mechanisms for autonomous interfacial ``micro''-robots \cite{chung2009electrowetting,surferbot}. 

We here introduce capillary surfers: highly-tunable solid particles that self-propel on a vibrating liquid surface due to the asymmetric radiation pressure of their self-generated surface waves. Two surfers interact via these surface waves and self-organize into multiple interaction modes, while multiple surfers exhibit collective behaviors. Generally, our results suggest that capillary surfers hold promise as a platform that bridges the gap between dissipation- and inertia-dominated active systems. 

\begin{figure}
    \centering
    \includegraphics[width=0.9\columnwidth]{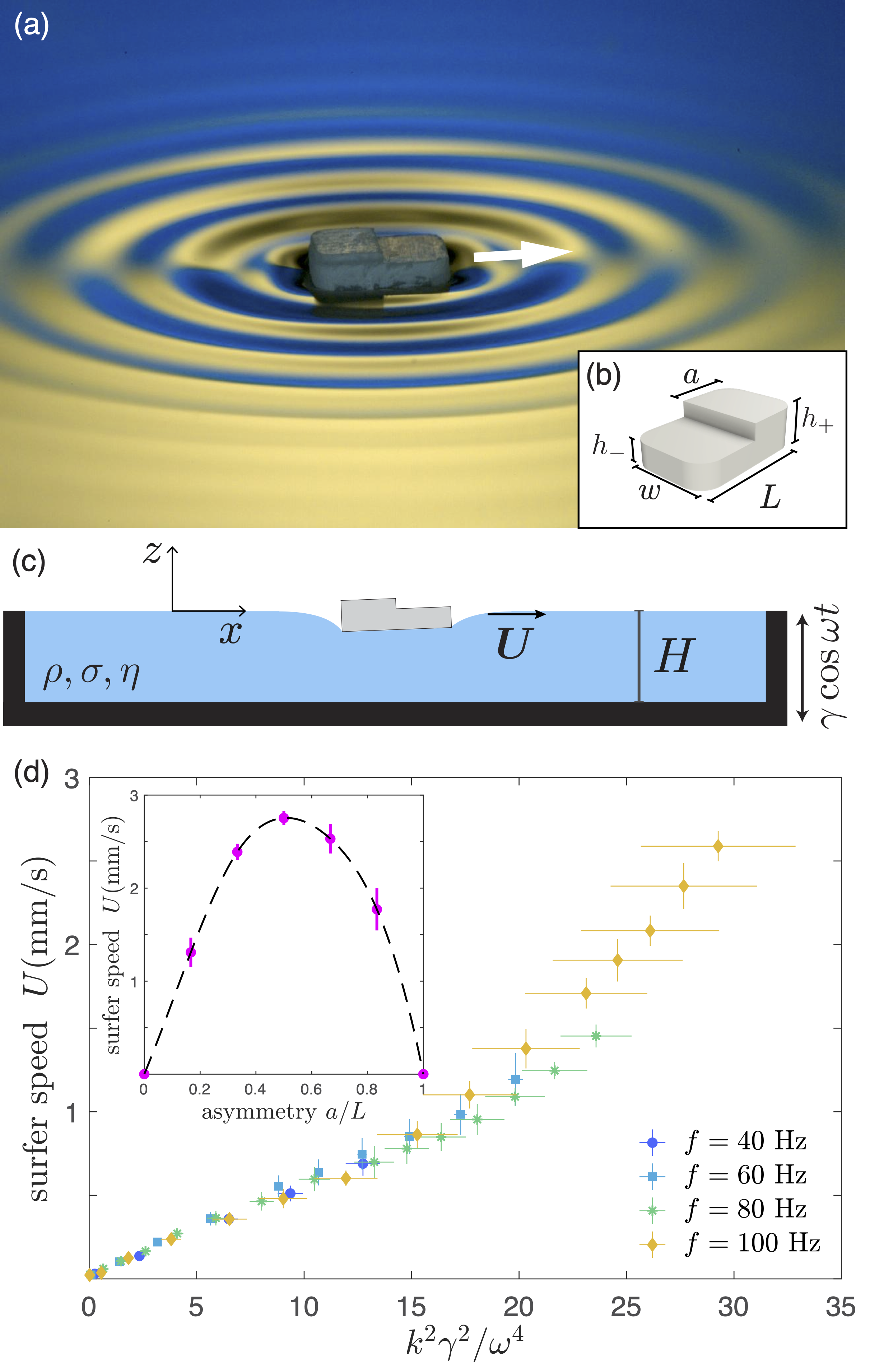}
    \caption{A capillary surfer self-propels on a fluid interface due to its self-generated waves (Supplementary Video 1). (a) Oblique wave field visualization, in which colors are obtained from the distorted reflection of a yellow and blue background on the fluid surface. (b) Surfer geometry. (c) Side view schematic of the experimental setup (not to scale). The fluid has density $\rho$, surface tension $\sigma$, dynamic viscosity $\eta$ and depth $H$. (d) Dependence of the surfer speed $U$ on the forcing frequency $f$ and forcing acceleration $\gamma$, as collapsed by \textcolor{black}{the non-dimensional propulsive force scaling in} Eq.~\eqref{eq:U}, with $L=4.20\pm0.03$~mm, $w=2.70\pm0.03$~mm, $a=L/2$. Inset: dependence of $U$ on the asymmetry $a/L$ for $f=100$~Hz, $\gamma=3.3$~g, $L=6.44\pm0.02$~mm, $w=4.13\pm0.02$~mm.}
    \label{fig:1}
\end{figure}

\textcolor{black}{Millimetric surfers (Fig.~\ref{fig:1}(b)) with width $w=1.75$--7.09~mm and length $L=2.78$--10.85~mm were manufactured out of white polytetrafluoroethylene (PTFE) sheets with density 2.2 g/cm$^3$, which are chemically hydrophobic. An adhesive PTFE sheet with thickness $0.40$~mm was attached to the upper surface of a PTFE sheet with thickness $0.82$~mm. A laser cutter (Universal Laser Systems, VLS 4.60) was then used to cut out rectangular profiles with rounded corners (with radius equal to $0.17L$) to avoid sharp corners along the contact line, which were observed to reduce the reproducibility of the surfer motion due to irregular wetting. In order to introduce an asymmetry in the surfer distribution of mass, a laser engraved line was used as a guide to cut out a portion of the upper adhered layer using a fine razor. Taking inspiration from marine terminology, in the following we refer to the front and back of the surfer as the ``bow'' and ``stern'', respectively. The height of the surfer's stern and bow were $h_+ = 1.22$~mm and $h_- = 0.82$~mm, respectively. The center of mass of the surfers was thus offset with respect to their in-plane geometric center.
Surfers were gently deposited on a bath of water-glycerol mixture with density $\rho=1.1756\pm0.0003$~g/cm$^3$, viscosity $\eta=0.0197\pm0.0005$~Pa$\cdot$s, surface tension $\sigma = 66.4\pm0.5$~mN/m and depth $H=5.73\pm0.06$~mm,} and supported at the liquid-air interface by virtue of the equilibrium between their weight, hydrostatic forces and surface tension. The contact line of the bath was pinned to the surfer's base perimeter. As a result of their mass asymmetry, surfers were slightly tilted in equilibrium (Fig. \ref{fig:1}(c)) and the deformation of the interface varied along their perimeter. \textcolor{black}{For the parameters explored here, the surfer's tilt during oscillation was substantially smaller than the tilt of asymmetric self-propelled granular particles on a vibrating plate \cite{dorbolo2005dynamics}. With a pinned contact line and symmetric contact area, mass asymmetry was a requirement to obtain self-propulsion, yet propulsion can be realized with virtually any surfer shape.}
The liquid bath was vertically driven by an electromagnetic shaker with acceleration $\Gamma(t) = \gamma \cos \omega t$, with $\omega = 2 \pi f$ and $f$ the forcing frequency in the range 40--100~Hz. \textcolor{black}{All experiments were performed below the Faraday threshold, the critical vibration amplitude above which subharmonic standing waves form spontaneously at the free surface \cite{faraday1831forms}. Steady and unsteady Faraday wave fields have been previously shown to result in passive particle migration \cite{yang2019passive} and redistribution \cite{falkovich2005floater,sanli2014antinode}. A monochrome USB camera (Allied Vision, Mako) with a macro lens for video acquisition was mounted above the bath and normal to its surface. In order to increase the contrast of the video recordings, the bath's base was constructed of a black acrylic plate. The shaker system was placed inside an acrylic box to isolate the bath from ambient air currents and contaminants.} More details on experimental methods and procedures are available in Supplemental Material~\cite{supmat}.

As soon as the bath was set into vibration, a surfer generated propagating surface waves as a result of the relative vertical motion between the surfer and bath, a consequence of the {\color{black} difference in inertia between the surfer and the liquid}. Correspondingly, the surfer moved along its long axis in the direction of its thinner half (Fig.~\ref{fig:1}(a,c)). \textcolor{black}{The contact line remained pinned to the surfer's base perimeter at all times. Due to the frequencies considered, surface waves were in the capillary regime with wavelength $\lambda=2\pi/k$ given by the dispersion relation $\omega^2=\sigma k^3/\rho$ in the deep-water limit $k H\gg 1$. In the parameter regime explored, $\lambda = 3.3 - 5.2$~mm, which is comparable to the surfer size.} In the absence of external perturbations or manufacturing imperfections, the surfer moved with constant velocity $\boldsymbol{U}$ along a rectilinear trajectory. We proceeded by analyzing the motion of a single surfer as a function of the forcing parameters, the surfer asymmetry, and its size (Fig.~\ref{fig:1}(d) and Supplementary Figs.~S1 and S2 in \cite{supmat}). For a given surfer size and asymmetry and a fixed forcing frequency $f$, the surfer speed increases with the forcing acceleration $\gamma$. For a fixed acceleration amplitude, the speed decreases with frequency. {\color{black} In all cases, the surfers move significantly slower than the phase speed of their self-generated propagating waves.}

\begin{figure*}
    \centering
    \includegraphics[width=\textwidth]{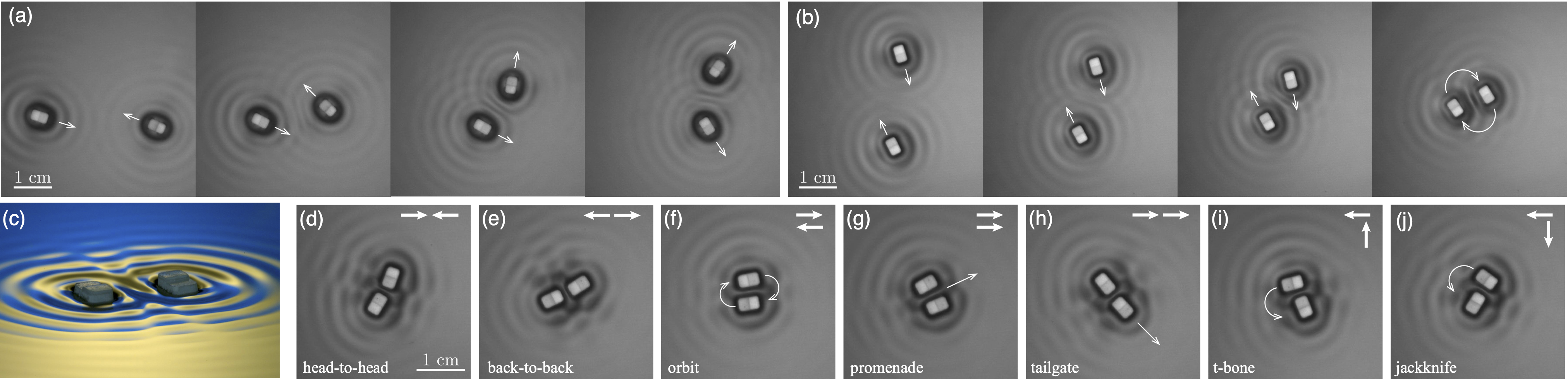}
    \caption{\textcolor{black}{Interaction of surfer pairs (see also Supplementary Video 2). (a) A collision yielding scattering. (b) A collision yielding a $n=2$ orbit. (c) Oblique view of two orbiting surfers. (d--j) Bound modes} observed in experiment with $n=1$ for $f = 100$ Hz and $\gamma=3.3$ g. (d) Head-to-head, (e) back-to-back, (f) orbit, (g) promenade, (h) tailgate, (i) t-bone, (f) jackknife. In experiments, $L=4.20\pm0.03$~mm, $w=2.70\pm0.03$~mm, $a=L/2$.} 
    \label{fig:2}
\end{figure*}

In order to rationalize this dependence on the driving parameters, we consider the wave radiation stress generated by the surfer. The radiation stress $S$ of surface waves may be defined as ``the excess flow of momentum due to the presence of the waves'' \cite{longuet1964radiation} and has the form $S = (3/4)\sigma A^2 k^2$ for capillary waves of amplitude $A$ and wavenumber $k$. 
Experimental measurements of the dependence of the surfer speed on its asymmetry show that the speed is maximized when the displacement of the center of mass relative to the in-plane geometric center is maximized, which corresponds to the highest difference in the equilibrium deformation of the liquid surface between stern and bow (inset of Fig.~\ref{fig:1}(d)). The radiation stress generated by the surfer thus exhibits a fore-aft asymmetry: waves of larger amplitude $A_+$ are generated at the stern, where the effective mass is larger, while waves of smaller amplitude $A_-$ are generated at the bow. As a result, the surfer experiences a net propulsive force
$F_p = (3/4)\sigma k^2 w(A_+^2-A_-^2)$, where $w$ is the surfer width. \textcolor{black}{ We assume that the wave amplitudes $A_{\pm}$ are proportional to the bath forcing amplitude $\gamma/\omega^2$ and define a non-dimensional propulsion force $F_p^*=F_p/(\sigma w)$ that thus scales as
\begin{align}
F_p^*\sim\left(\frac{k \gamma}{\omega^2}\right)^2.
\label{eq:U}
\end{align}
Figure~\ref{fig:1}(d) shows that the experimental speed measurements for a single surfer over a range of driving amplitudes and frequencies can be collapsed along a single curve when plotted as a function of this non-dimensional propulsion force (see raw data in Supplementary Fig.~S1 \cite{supmat}), confirming the hypothesis of wave radiation stress as the mechanism underlying surfer propulsion. 
 Additionally, for small speeds, this curve follows a linear trend which suggests a corresponding linear (viscous) resistance.}

Surfers may interact with each other through the propagating waves they generate on the fluid interface. \textcolor{black}{Specifically, when two free surfers come within a few capillary wavelengths of each other, they deviate markedly from their rectilinear motion. This wave-mediated interaction has two possible outcomes. In a scattering process the surfers repel each other and ultimately recover rectilinear motions in new directions (Fig.~\ref{fig:2}(a)). Alternatively, the surfers can enter a stable bound mode after a brief transient period (Fig.~\ref{fig:2}(b)). In general, the outcome of the interaction depends on the initial condition of the surfers' relative motion. The number of bound modes varies with forcing frequency and amplitude. When the system was driven at $f=100$~Hz and $\gamma=3.3$~g ({\itshape{i.e.}} close to but below the Faraday instability threshold $\gamma_F=3.80\pm0.05$~g) two surfers of equal size and speed with length $L=4.20\pm0.03$~mm, width $w=2.70\pm0.03$~mm and $a=L/2$ exhibited the maximum number of bound modes. Variation of the experimental parameters, including liquid viscosity, generally resulted in a reduction of the number of observable modes. With these parameters,} we observed up to seven qualitatively distinct bound modes (Fig.~\ref{fig:2}(d--j) and Supplementary Video 2): head-to-head (d), back-to-back (e), orbiting (f), promenade (g), tailgating (h), t-bone (i) and jackknife (j). In the head-to-head (back-to-back) mode the two surfer bows (sterns) face each other (Fig.~\ref{fig:2}(d,e)). While these modes are static, the remaining five modes are dynamic. In the orbiting mode, the two surfers orbit around the system's center of mass (Fig.~\ref{fig:2}(f)). In the promenade mode, they proceed side by side with constant speed along a rectilinear trajectory (Fig.~\ref{fig:2}(g)). In the tailgating mode, the two surfers are aligned along their major axis, with the bow of one surfer pointing toward the stern of the other, and they move with constant speed along a rectilinear trajectory (Fig.~\ref{fig:2}(h)). In the t-bone mode, the two major axes are perpendicular to each other and the bow of one surfer points toward the stern of the other, while they both execute a circular trajectory (Fig.~\ref{fig:2}(i)). The jackknife mode has a similar configuration except the two sterns {\textcolor{black}{are close to} each other (Fig.~\ref{fig:2}(j)). \textcolor{black}{This rich catalog of bound modes shares many similarities with recent numerical predictions of pairwise bound modes for fully immersed two-sphere intermediate Reynolds number swimmers \cite{dombrowski2022pairwise}. In our system, bound modes can be obtained by either varying the initial conditions of a two-surfer impact or by varying initial positions and orientations of two surfers manually positioned near one another. The back-to-back, tailgaiting and jackknife modes could not be realized in an impact event.} Note that should the vibration be eliminated, the surfers cease to propel and immediately collapse in towards each other under the influence of capillary attraction \cite{ho2019direct}.

\begin{figure}
    \centering
    \includegraphics[width=1\columnwidth]{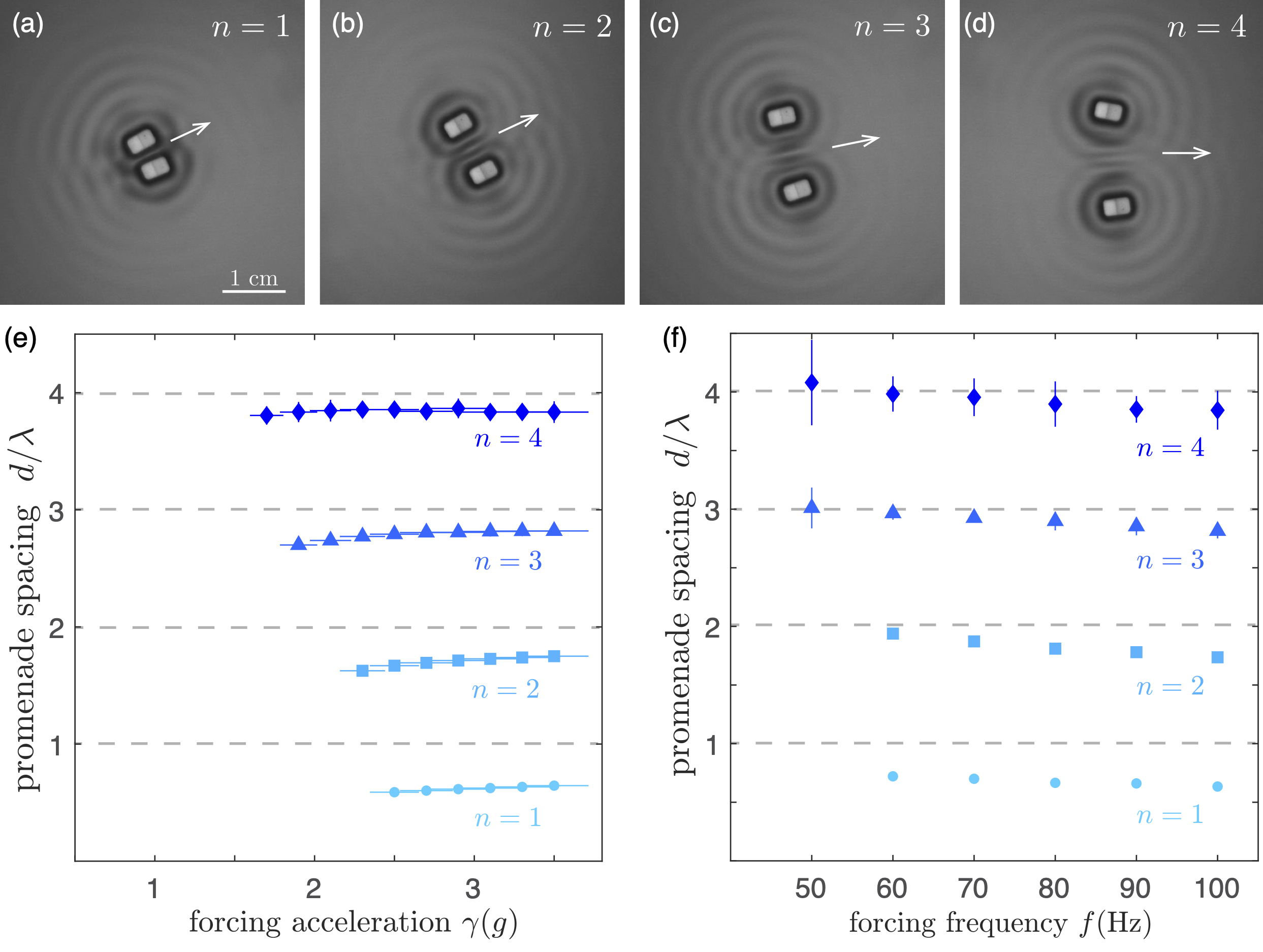}
    \caption{Promenade modes. In (a)--(d), $\gamma = 3.3$~g and $f = 100$~Hz. (e) Dependence of the dimensionless inter-surfer spacing $d/\lambda$ on the forcing acceleration $\gamma$ for fixed forcing frequency $f=100$~Hz. (f) Dependence of $d/\lambda$ on $f$ just below the Faraday instability threshold. $f$ ranges from 50--100 Hz in increments of 10 Hz, and the corresponding values of $\gamma/g$ are 1.1, 1.5, 2.0, 2.3, 3.0 and 3.3. $d$ is defined as the distance between the surfers. $L=4.20\pm0.03$~mm, $w=2.70\pm0.03$~mm and $a=L/2$.}
    \label{fig:3}
\end{figure}
In each mode the two surfers exhibit discrete equilibrium spacings $d \approx n \lambda$ where $n=1,2,\dots$. 
The maximum order $n_{\rm max}$ {\color{black}is different for each mode and} depends on the properties of the liquid, vibration, and surfers. {\color{black}$n_{\rm max}$ and the individual surfer speeds for each mode are available in the Supplemental Material~\cite{supmat}}. 
We focused on the promenade mode as in our experiments it exhibited the largest $n_{\rm max}=4$ (Fig.~\ref{fig:3}). First, we fixed the forcing frequency and measured the promenade spacing for increasing forcing acceleration. The spacing $d$ increases very slightly with forcing acceleration $\gamma$ (Fig.~\ref{fig:3}(e)), but is significantly more sensitive to the forcing frequency $f$. The normalized spacing $d/\lambda$ is approximately independent of $f$, showing that the equilibrium spacings are quantized by the capillary wavelength in the range of forcing frequencies explored (Fig.~\ref{fig:3}(f)). \textcolor{black}{The promenade speed increased with $n$ but was always lower than the speed of a single surfer.}

Many-body experiments show that capillary surfers have potential as constituents of a novel active system (Fig.~\ref{fig:4} and Supplementary Video 3). For instance, a many-body promenade mode (Fig.~\ref{fig:4}(a)) and a super-orbiting state of eight surfers (Fig.~\ref{fig:4}(b)) have been observed in experiments. \textcolor{black}{Multiple surfers can also arrange in static lattices (Fig.~\ref{fig:4}(c)) or exhibit exotic roto-translating bound states (Fig.~\ref{fig:4}(d)).}
While there have been extensive studies on overdamped active systems~\cite{marchetti2013hydrodynamics}, mediated by viscous hydrodynamic forces that decay monotonically with distance, a collection of surfers has the peculiar feature of being characterized by wave-mediated interactions, which results in long-range spatially-oscillatory forces defined by alternating regions of attraction and repulsion. This feature is a consequence of fluid inertia, and responsible for the multistability of a discrete set of interaction states, as documented here for the promenade mode (Fig.~\ref{fig:3}). 

\begin{figure}
    \centering
    \includegraphics[width=1\columnwidth]{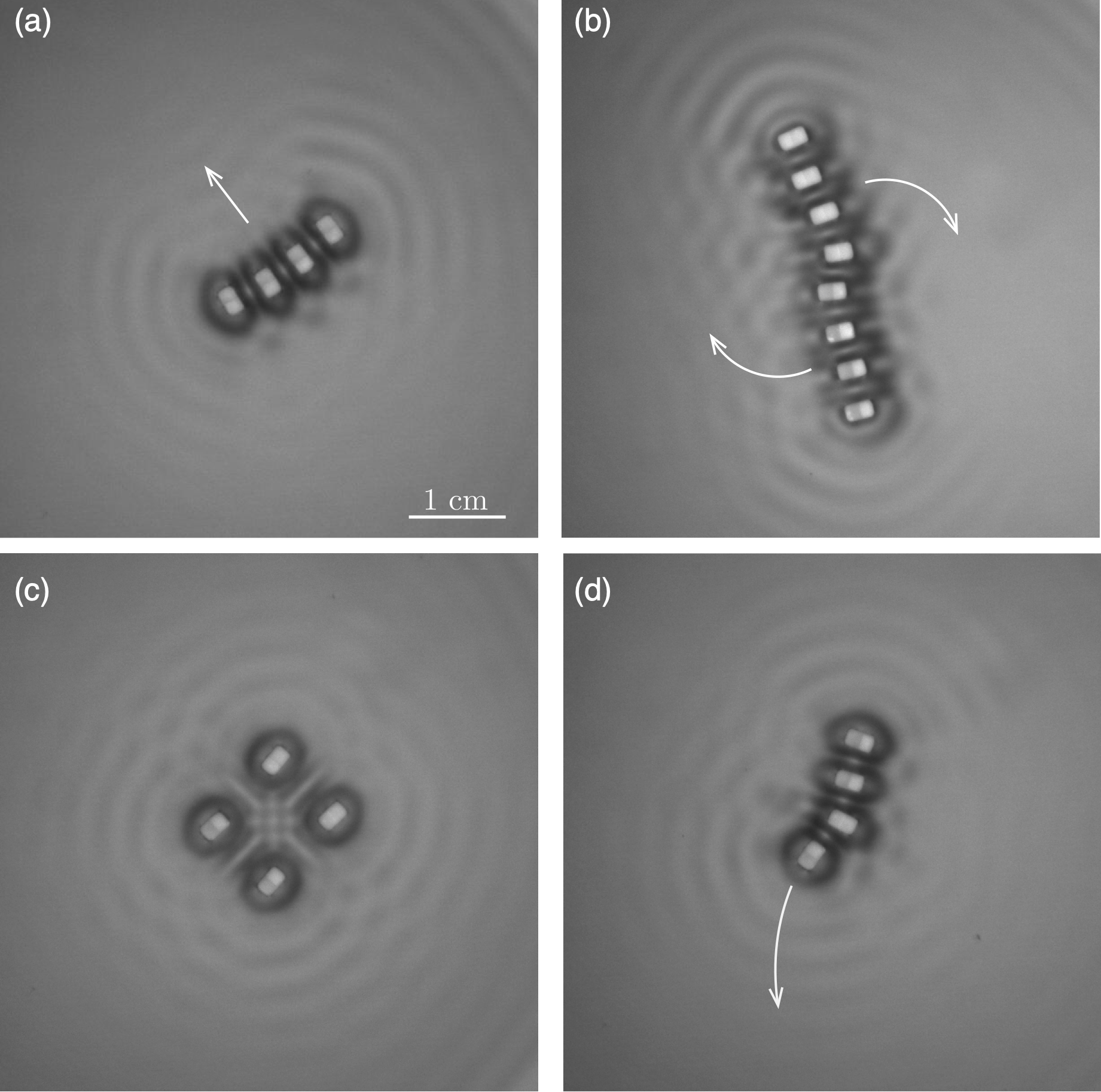}
    \caption{Examples of collective behavior of surfers (Supplementary Video 3). (a) Promenade mode of four surfers. (b) Super-orbiting mode of eight surfers. (c) Lattice mode of four surfers. (d) Composite mode of four surfers. Experimental parameters: $f=100$~Hz, $\gamma=3.3$~g, $L=2.78\pm0.01$~mm, $w=1.75\pm0.01$~mm, $a=L/2$.}
    \label{fig:4}
\end{figure}

Collective behaviors at a fluid interface have been observed with camphor boats \cite{suematsu2010collective} and walking droplets \cite{saenz2018spin} previously,
but only within a relatively narrow parameter space defined by the physical constraints of these systems. The two-surfer interaction landscape (Fig.~\ref{fig:2}) already far exceeds those documented for the aforementioned systems.  Ultimately, the surfer system is a \textcolor{black}{largely} tunable and accessible experimental platform that has the potential to fill the gap between active systems at low and high Reynolds numbers: the surfers used in the present work self-propel at intermediate Reynolds number $\text{Re}=\rho UL/\eta \sim O(1)$, where both fluid inertia and viscous forces are relevant. Recent reviews \cite{bechinger2016active,klotsa2019above} underline the need for experimental platforms and theoretical frameworks to explore active systems in this intermediate regime, which has received far less attention than the purely viscous and inertial limits. It is anticipated that such systems could exhibit novel classes of self-organising behaviors, such as soliton-like waves, shock-like phenomena, new flocking states (e.g. Fig.~\ref{fig:4}) and non-equilibrium phase transitions \cite{bechinger2016active,klotsa2019above}.

\begin{acknowledgements}
AUO acknowledges support from the Simons Foundation (Collaboration Grant for Mathematicians, Award No. 587006) and NSF DMS-210883. GP acknowledges the CNRS Momentum program for its support.  DMH acknowledges support from the Office of Naval Research (ONR N00014-21-1-2816) and the Brown Undergraduate Teaching and Research Award. Additionally, the authors thank Prof. Roberto Zenit for use of the conical rheometer. 
\end{acknowledgements}

\bibliographystyle{apsrev4-2}
\bibliography{biblio}

\begin{thebibliography}{65}%
\makeatletter
\providecommand \@ifxundefined [1]{%
 \@ifx{#1\undefined}
}%
\providecommand \@ifnum [1]{%
 \ifnum #1\expandafter \@firstoftwo
 \else \expandafter \@secondoftwo
 \fi
}%
\providecommand \@ifx [1]{%
 \ifx #1\expandafter \@firstoftwo
 \else \expandafter \@secondoftwo
 \fi
}%
\providecommand \natexlab [1]{#1}%
\providecommand \enquote  [1]{``#1''}%
\providecommand \bibnamefont  [1]{#1}%
\providecommand \bibfnamefont [1]{#1}%
\providecommand \citenamefont [1]{#1}%
\providecommand \href@noop [0]{\@secondoftwo}%
\providecommand \href [0]{\begingroup \@sanitize@url \@href}%
\providecommand \@href[1]{\@@startlink{#1}\@@href}%
\providecommand \@@href[1]{\endgroup#1\@@endlink}%
\providecommand \@sanitize@url [0]{\catcode `\\12\catcode `\$12\catcode
  `\&12\catcode `\#12\catcode `\^12\catcode `\_12\catcode `\%12\relax}%
\providecommand \@@startlink[1]{}%
\providecommand \@@endlink[0]{}%
\providecommand \url  [0]{\begingroup\@sanitize@url \@url }%
\providecommand \@url [1]{\endgroup\@href {#1}{\urlprefix }}%
\providecommand \urlprefix  [0]{URL }%
\providecommand \Eprint [0]{\href }%
\providecommand \doibase [0]{https://doi.org/}%
\providecommand \selectlanguage [0]{\@gobble}%
\providecommand \bibinfo  [0]{\@secondoftwo}%
\providecommand \bibfield  [0]{\@secondoftwo}%
\providecommand \translation [1]{[#1]}%
\providecommand \BibitemOpen [0]{}%
\providecommand \bibitemStop [0]{}%
\providecommand \bibitemNoStop [0]{.\EOS\space}%
\providecommand \EOS [0]{\spacefactor3000\relax}%
\providecommand \BibitemShut  [1]{\csname bibitem#1\endcsname}%
\let\auto@bib@innerbib\@empty
\bibitem [{\citenamefont {Ramaswamy}(2010)}]{ramaswamy2010mechanics}%
  \BibitemOpen
  \bibfield  {author} {\bibinfo {author} {\bibfnamefont {S.}~\bibnamefont
  {Ramaswamy}},\ }\href@noop {} {\bibfield  {journal} {\bibinfo  {journal}
  {Annu. Rev. Condens. Matter Phys.}\ }\textbf {\bibinfo {volume} {1}},\
  \bibinfo {pages} {323} (\bibinfo {year} {2010})}\BibitemShut {NoStop}%
\bibitem [{\citenamefont {Marchetti}\ \emph {et~al.}(2013)\citenamefont
  {Marchetti}, \citenamefont {Joanny}, \citenamefont {Ramaswamy}, \citenamefont
  {Liverpool}, \citenamefont {Prost}, \citenamefont {Rao},\ and\ \citenamefont
  {Simha}}]{marchetti2013hydrodynamics}%
  \BibitemOpen
  \bibfield  {author} {\bibinfo {author} {\bibfnamefont {M.~C.}\ \bibnamefont
  {Marchetti}}, \bibinfo {author} {\bibfnamefont {J.-F.}\ \bibnamefont
  {Joanny}}, \bibinfo {author} {\bibfnamefont {S.}~\bibnamefont {Ramaswamy}},
  \bibinfo {author} {\bibfnamefont {T.~B.}\ \bibnamefont {Liverpool}}, \bibinfo
  {author} {\bibfnamefont {J.}~\bibnamefont {Prost}}, \bibinfo {author}
  {\bibfnamefont {M.}~\bibnamefont {Rao}},\ and\ \bibinfo {author}
  {\bibfnamefont {R.~A.}\ \bibnamefont {Simha}},\ }\href@noop {} {\bibfield
  {journal} {\bibinfo  {journal} {Rev. Mod. Phys.}\ }\textbf {\bibinfo {volume}
  {85}},\ \bibinfo {pages} {1143} (\bibinfo {year} {2013})}\BibitemShut
  {NoStop}%
\bibitem [{\citenamefont {Saintillan}\ and\ \citenamefont
  {Shelley}(2007)}]{saintillan2007orientational}%
  \BibitemOpen
  \bibfield  {author} {\bibinfo {author} {\bibfnamefont {D.}~\bibnamefont
  {Saintillan}}\ and\ \bibinfo {author} {\bibfnamefont {M.~J.}\ \bibnamefont
  {Shelley}},\ }\href@noop {} {\bibfield  {journal} {\bibinfo  {journal} {Phys.
  Rev. Lett.}\ }\textbf {\bibinfo {volume} {99}},\ \bibinfo {pages} {058102}
  (\bibinfo {year} {2007})}\BibitemShut {NoStop}%
\bibitem [{\citenamefont {Sokolov}\ and\ \citenamefont
  {Aranson}(2009)}]{Sokolov_viscosity}%
  \BibitemOpen
  \bibfield  {author} {\bibinfo {author} {\bibfnamefont {A.}~\bibnamefont
  {Sokolov}}\ and\ \bibinfo {author} {\bibfnamefont {I.~S.}\ \bibnamefont
  {Aranson}},\ }\href@noop {} {\bibfield  {journal} {\bibinfo  {journal} {Phys.
  Rev. Lett.}\ }\textbf {\bibinfo {volume} {103}},\ \bibinfo {pages} {148101}
  (\bibinfo {year} {2009})}\BibitemShut {NoStop}%
\bibitem [{\citenamefont {Sanchez}\ \emph {et~al.}(2012)\citenamefont
  {Sanchez}, \citenamefont {Chen}, \citenamefont {DeCamp}, \citenamefont
  {Heymann},\ and\ \citenamefont {Dogic}}]{sanchez2012spontaneous}%
  \BibitemOpen
  \bibfield  {author} {\bibinfo {author} {\bibfnamefont {T.}~\bibnamefont
  {Sanchez}}, \bibinfo {author} {\bibfnamefont {D.~T.~N.}\ \bibnamefont
  {Chen}}, \bibinfo {author} {\bibfnamefont {S.~J.}\ \bibnamefont {DeCamp}},
  \bibinfo {author} {\bibfnamefont {M.}~\bibnamefont {Heymann}},\ and\ \bibinfo
  {author} {\bibfnamefont {Z.}~\bibnamefont {Dogic}},\ }\href@noop {}
  {\bibfield  {journal} {\bibinfo  {journal} {Nature}\ }\textbf {\bibinfo
  {volume} {491}},\ \bibinfo {pages} {431} (\bibinfo {year}
  {2012})}\BibitemShut {NoStop}%
\bibitem [{\citenamefont {Wensink}\ \emph {et~al.}(2012)\citenamefont
  {Wensink}, \citenamefont {Dunkel}, \citenamefont {Heidenreich}, \citenamefont
  {Drescher}, \citenamefont {Goldstein}, \citenamefont {L\"{o}wen},\ and\
  \citenamefont {Yeomans}}]{Goldstein_Turbulence}%
  \BibitemOpen
  \bibfield  {author} {\bibinfo {author} {\bibfnamefont {H.~H.}\ \bibnamefont
  {Wensink}}, \bibinfo {author} {\bibfnamefont {J.}~\bibnamefont {Dunkel}},
  \bibinfo {author} {\bibfnamefont {S.}~\bibnamefont {Heidenreich}}, \bibinfo
  {author} {\bibfnamefont {K.}~\bibnamefont {Drescher}}, \bibinfo {author}
  {\bibfnamefont {R.~E.}\ \bibnamefont {Goldstein}}, \bibinfo {author}
  {\bibfnamefont {H.}~\bibnamefont {L\"{o}wen}},\ and\ \bibinfo {author}
  {\bibfnamefont {J.~M.}\ \bibnamefont {Yeomans}},\ }\href@noop {} {\bibfield
  {journal} {\bibinfo  {journal} {Proc. Natl. Acad. Sci.}\ }\textbf {\bibinfo
  {volume} {109}},\ \bibinfo {pages} {14308} (\bibinfo {year}
  {2012})}\BibitemShut {NoStop}%
\bibitem [{\citenamefont {Palacci}\ \emph {et~al.}(2013)\citenamefont
  {Palacci}, \citenamefont {Sacanna}, \citenamefont {Steinberg}, \citenamefont
  {Pine},\ and\ \citenamefont {Chaikin}}]{Palacci_Light}%
  \BibitemOpen
  \bibfield  {author} {\bibinfo {author} {\bibfnamefont {J.}~\bibnamefont
  {Palacci}}, \bibinfo {author} {\bibfnamefont {S.}~\bibnamefont {Sacanna}},
  \bibinfo {author} {\bibfnamefont {A.~P.}\ \bibnamefont {Steinberg}}, \bibinfo
  {author} {\bibfnamefont {D.~J.}\ \bibnamefont {Pine}},\ and\ \bibinfo
  {author} {\bibfnamefont {P.~M.}\ \bibnamefont {Chaikin}},\ }\href@noop {}
  {\bibfield  {journal} {\bibinfo  {journal} {Science}\ }\textbf {\bibinfo
  {volume} {339}},\ \bibinfo {pages} {936} (\bibinfo {year}
  {2013})}\BibitemShut {NoStop}%
\bibitem [{\citenamefont {Golestanian}(2012)}]{golestanian2012collective}%
  \BibitemOpen
  \bibfield  {author} {\bibinfo {author} {\bibfnamefont {R.}~\bibnamefont
  {Golestanian}},\ }\href@noop {} {\bibfield  {journal} {\bibinfo  {journal}
  {Phys. Rev. Lett.}\ }\textbf {\bibinfo {volume} {108}},\ \bibinfo {pages}
  {038303} (\bibinfo {year} {2012})}\BibitemShut {NoStop}%
\bibitem [{\citenamefont {Solon}\ \emph {et~al.}(2015)\citenamefont {Solon},
  \citenamefont {Stenhammar}, \citenamefont {Wittkowski}, \citenamefont
  {Kardar}, \citenamefont {Kafri}, \citenamefont {Cates},\ and\ \citenamefont
  {Tailleur}}]{Solon2015}%
  \BibitemOpen
  \bibfield  {author} {\bibinfo {author} {\bibfnamefont {A.~P.}\ \bibnamefont
  {Solon}}, \bibinfo {author} {\bibfnamefont {J.}~\bibnamefont {Stenhammar}},
  \bibinfo {author} {\bibfnamefont {R.}~\bibnamefont {Wittkowski}}, \bibinfo
  {author} {\bibfnamefont {M.}~\bibnamefont {Kardar}}, \bibinfo {author}
  {\bibfnamefont {Y.}~\bibnamefont {Kafri}}, \bibinfo {author} {\bibfnamefont
  {M.~E.}\ \bibnamefont {Cates}},\ and\ \bibinfo {author} {\bibfnamefont
  {J.}~\bibnamefont {Tailleur}},\ }\href
  {https://doi.org/10.1103/PhysRevLett.114.198301} {\bibfield  {journal}
  {\bibinfo  {journal} {Phys. Rev. Lett.}\ }\textbf {\bibinfo {volume} {114}},\
  \bibinfo {pages} {198301} (\bibinfo {year} {2015})}\BibitemShut {NoStop}%
\bibitem [{\citenamefont {Brosseau}\ \emph {et~al.}(2019)\citenamefont
  {Brosseau}, \citenamefont {Usabiaga}, \citenamefont {Lushi}, \citenamefont
  {Wu}, \citenamefont {Ristroph}, \citenamefont {Zhang}, \citenamefont {Ward},\
  and\ \citenamefont {Shelley}}]{Brosseau2019}%
  \BibitemOpen
  \bibfield  {author} {\bibinfo {author} {\bibfnamefont {Q.}~\bibnamefont
  {Brosseau}}, \bibinfo {author} {\bibfnamefont {F.~B.}\ \bibnamefont
  {Usabiaga}}, \bibinfo {author} {\bibfnamefont {E.}~\bibnamefont {Lushi}},
  \bibinfo {author} {\bibfnamefont {Y.}~\bibnamefont {Wu}}, \bibinfo {author}
  {\bibfnamefont {L.}~\bibnamefont {Ristroph}}, \bibinfo {author}
  {\bibfnamefont {J.}~\bibnamefont {Zhang}}, \bibinfo {author} {\bibfnamefont
  {M.}~\bibnamefont {Ward}},\ and\ \bibinfo {author} {\bibfnamefont {M.~J.}\
  \bibnamefont {Shelley}},\ }\href
  {https://doi.org/10.1103/PhysRevLett.123.178004} {\bibfield  {journal}
  {\bibinfo  {journal} {Phys. Rev. Lett.}\ }\textbf {\bibinfo {volume} {123}},\
  \bibinfo {pages} {178004} (\bibinfo {year} {2019})}\BibitemShut {NoStop}%
\bibitem [{\citenamefont {Huber}\ \emph {et~al.}(2018)\citenamefont {Huber},
  \citenamefont {Suzuki}, \citenamefont {Kr{\"u}ger}, \citenamefont {Frey},\
  and\ \citenamefont {Bausch}}]{huber2018emergence}%
  \BibitemOpen
  \bibfield  {author} {\bibinfo {author} {\bibfnamefont {L.}~\bibnamefont
  {Huber}}, \bibinfo {author} {\bibfnamefont {R.}~\bibnamefont {Suzuki}},
  \bibinfo {author} {\bibfnamefont {T.}~\bibnamefont {Kr{\"u}ger}}, \bibinfo
  {author} {\bibfnamefont {E.}~\bibnamefont {Frey}},\ and\ \bibinfo {author}
  {\bibfnamefont {A.}~\bibnamefont {Bausch}},\ }\href@noop {} {\bibfield
  {journal} {\bibinfo  {journal} {Science}\ }\textbf {\bibinfo {volume}
  {361}},\ \bibinfo {pages} {255} (\bibinfo {year} {2018})}\BibitemShut
  {NoStop}%
\bibitem [{\citenamefont {Doostmohammadi}\ \emph {et~al.}(2018)\citenamefont
  {Doostmohammadi}, \citenamefont {Ign\'{e}s-Mullol}, \citenamefont {Yeomans},\
  and\ \citenamefont {Sagu\'{e}s}}]{doostmohammadi2018nematic}%
  \BibitemOpen
  \bibfield  {author} {\bibinfo {author} {\bibfnamefont {A.}~\bibnamefont
  {Doostmohammadi}}, \bibinfo {author} {\bibfnamefont {J.}~\bibnamefont
  {Ign\'{e}s-Mullol}}, \bibinfo {author} {\bibfnamefont {J.~M.}\ \bibnamefont
  {Yeomans}},\ and\ \bibinfo {author} {\bibfnamefont {F.}~\bibnamefont
  {Sagu\'{e}s}},\ }\href@noop {} {\bibfield  {journal} {\bibinfo  {journal}
  {Nat. Commun.}\ }\textbf {\bibinfo {volume} {9}},\ \bibinfo {pages} {3246}
  (\bibinfo {year} {2018})}\BibitemShut {NoStop}%
\bibitem [{\citenamefont {Soni}\ \emph {et~al.}(2019)\citenamefont {Soni},
  \citenamefont {Bililign}, \citenamefont {Magkiriadou}, \citenamefont
  {Sacanna}, \citenamefont {Bartolo}, \citenamefont {Shelley},\ and\
  \citenamefont {Irvine}}]{Soni2019}%
  \BibitemOpen
  \bibfield  {author} {\bibinfo {author} {\bibfnamefont {V.}~\bibnamefont
  {Soni}}, \bibinfo {author} {\bibfnamefont {E.~S.}\ \bibnamefont {Bililign}},
  \bibinfo {author} {\bibfnamefont {S.}~\bibnamefont {Magkiriadou}}, \bibinfo
  {author} {\bibfnamefont {S.}~\bibnamefont {Sacanna}}, \bibinfo {author}
  {\bibfnamefont {D.}~\bibnamefont {Bartolo}}, \bibinfo {author} {\bibfnamefont
  {M.~J.}\ \bibnamefont {Shelley}},\ and\ \bibinfo {author} {\bibfnamefont
  {W.~T.~M.}\ \bibnamefont {Irvine}},\ }\href@noop {} {\bibfield  {journal}
  {\bibinfo  {journal} {Nat. Phys.}\ }\textbf {\bibinfo {volume} {15}},\
  \bibinfo {pages} {1188–1194} (\bibinfo {year} {2019})}\BibitemShut
  {NoStop}%
\bibitem [{\citenamefont {Bricard}\ \emph {et~al.}(2013)\citenamefont
  {Bricard}, \citenamefont {Caussin}, \citenamefont {Desreumaux}, \citenamefont
  {Dauchot},\ and\ \citenamefont {Bartolo}}]{Bricard_Rollers}%
  \BibitemOpen
  \bibfield  {author} {\bibinfo {author} {\bibfnamefont {A.}~\bibnamefont
  {Bricard}}, \bibinfo {author} {\bibfnamefont {J.-B.}\ \bibnamefont
  {Caussin}}, \bibinfo {author} {\bibfnamefont {N.}~\bibnamefont {Desreumaux}},
  \bibinfo {author} {\bibfnamefont {O.}~\bibnamefont {Dauchot}},\ and\ \bibinfo
  {author} {\bibfnamefont {D.}~\bibnamefont {Bartolo}},\ }\href@noop {}
  {\bibfield  {journal} {\bibinfo  {journal} {Nature}\ }\textbf {\bibinfo
  {volume} {503}},\ \bibinfo {pages} {95} (\bibinfo {year} {2013})}\BibitemShut
  {NoStop}%
\bibitem [{\citenamefont {Driscoll}\ \emph {et~al.}(2016)\citenamefont
  {Driscoll}, \citenamefont {Delmotte}, \citenamefont {Youssef}, \citenamefont
  {Sacanna}, \citenamefont {Donev},\ and\ \citenamefont
  {Chaikin}}]{Driscoll_Rollers}%
  \BibitemOpen
  \bibfield  {author} {\bibinfo {author} {\bibfnamefont {M.}~\bibnamefont
  {Driscoll}}, \bibinfo {author} {\bibfnamefont {B.}~\bibnamefont {Delmotte}},
  \bibinfo {author} {\bibfnamefont {M.}~\bibnamefont {Youssef}}, \bibinfo
  {author} {\bibfnamefont {S.}~\bibnamefont {Sacanna}}, \bibinfo {author}
  {\bibfnamefont {A.}~\bibnamefont {Donev}},\ and\ \bibinfo {author}
  {\bibfnamefont {P.}~\bibnamefont {Chaikin}},\ }\href@noop {} {\bibfield
  {journal} {\bibinfo  {journal} {Nat. Phys.}\ }\textbf {\bibinfo {volume}
  {13}},\ \bibinfo {pages} {375} (\bibinfo {year} {2016})}\BibitemShut
  {NoStop}%
\bibitem [{\citenamefont {Yeo}\ \emph {et~al.}(2015)\citenamefont {Yeo},
  \citenamefont {Lushi},\ and\ \citenamefont {Vlahovska}}]{Yeo_Rollers}%
  \BibitemOpen
  \bibfield  {author} {\bibinfo {author} {\bibfnamefont {K.}~\bibnamefont
  {Yeo}}, \bibinfo {author} {\bibfnamefont {E.}~\bibnamefont {Lushi}},\ and\
  \bibinfo {author} {\bibfnamefont {P.~M.}\ \bibnamefont {Vlahovska}},\
  }\href@noop {} {\bibfield  {journal} {\bibinfo  {journal} {Phys. Rev. Lett.}\
  }\textbf {\bibinfo {volume} {114}},\ \bibinfo {pages} {188301} (\bibinfo
  {year} {2015})}\BibitemShut {NoStop}%
\bibitem [{\citenamefont {Lissaman}\ and\ \citenamefont
  {Shollenberger}(1970)}]{lissaman1970formation}%
  \BibitemOpen
  \bibfield  {author} {\bibinfo {author} {\bibfnamefont {P.~B.~S.}\
  \bibnamefont {Lissaman}}\ and\ \bibinfo {author} {\bibfnamefont {C.~A.}\
  \bibnamefont {Shollenberger}},\ }\href@noop {} {\bibfield  {journal}
  {\bibinfo  {journal} {Science}\ }\textbf {\bibinfo {volume} {168}} (\bibinfo
  {year} {1970})}\BibitemShut {NoStop}%
\bibitem [{\citenamefont {Weihs}(1973)}]{weihs1973hydromechanics}%
  \BibitemOpen
  \bibfield  {author} {\bibinfo {author} {\bibfnamefont {D.}~\bibnamefont
  {Weihs}},\ }\href {https://doi.org/10.1038/241290a0} {\bibfield  {journal}
  {\bibinfo  {journal} {Nature}\ }\textbf {\bibinfo {volume} {241}},\ \bibinfo
  {pages} {290} (\bibinfo {year} {1973})}\BibitemShut {NoStop}%
\bibitem [{\citenamefont {Attanasi}\ \emph {et~al.}(2014)\citenamefont
  {Attanasi}, \citenamefont {Cavagna}, \citenamefont {Castello}, \citenamefont
  {Giardina}, \citenamefont {Grigera}, \citenamefont {Jeli\'{c}}, \citenamefont
  {Melillo}, \citenamefont {Parisi}, \citenamefont {Pohl}, \citenamefont
  {Shen},\ and\ \citenamefont {Viale}}]{AttanasiNatPhys}%
  \BibitemOpen
  \bibfield  {author} {\bibinfo {author} {\bibfnamefont {A.}~\bibnamefont
  {Attanasi}}, \bibinfo {author} {\bibfnamefont {A.}~\bibnamefont {Cavagna}},
  \bibinfo {author} {\bibfnamefont {L.~D.}\ \bibnamefont {Castello}}, \bibinfo
  {author} {\bibfnamefont {I.}~\bibnamefont {Giardina}}, \bibinfo {author}
  {\bibfnamefont {T.~S.}\ \bibnamefont {Grigera}}, \bibinfo {author}
  {\bibfnamefont {A.}~\bibnamefont {Jeli\'{c}}}, \bibinfo {author}
  {\bibfnamefont {S.}~\bibnamefont {Melillo}}, \bibinfo {author} {\bibfnamefont
  {L.}~\bibnamefont {Parisi}}, \bibinfo {author} {\bibfnamefont
  {O.}~\bibnamefont {Pohl}}, \bibinfo {author} {\bibfnamefont {E.}~\bibnamefont
  {Shen}},\ and\ \bibinfo {author} {\bibfnamefont {M.}~\bibnamefont {Viale}},\
  }\href@noop {} {\bibfield  {journal} {\bibinfo  {journal} {Nat. Phys.}\
  }\textbf {\bibinfo {volume} {10}},\ \bibinfo {pages} {691} (\bibinfo {year}
  {2014})}\BibitemShut {NoStop}%
\bibitem [{\citenamefont {Becker}\ \emph {et~al.}(2015)\citenamefont {Becker},
  \citenamefont {Masoud}, \citenamefont {Newbolt}, \citenamefont {Shelley},\
  and\ \citenamefont {Ristroph}}]{becker2015hydrodynamic}%
  \BibitemOpen
  \bibfield  {author} {\bibinfo {author} {\bibfnamefont {A.~D.}\ \bibnamefont
  {Becker}}, \bibinfo {author} {\bibfnamefont {H.}~\bibnamefont {Masoud}},
  \bibinfo {author} {\bibfnamefont {J.~W.}\ \bibnamefont {Newbolt}}, \bibinfo
  {author} {\bibfnamefont {M.}~\bibnamefont {Shelley}},\ and\ \bibinfo {author}
  {\bibfnamefont {L.}~\bibnamefont {Ristroph}},\ }\href@noop {} {\bibfield
  {journal} {\bibinfo  {journal} {Nat. Commun.}\ }\textbf {\bibinfo {volume}
  {6}},\ \bibinfo {pages} {8514} (\bibinfo {year} {2015})}\BibitemShut
  {NoStop}%
\bibitem [{\citenamefont {Oza}\ \emph {et~al.}(2019)\citenamefont {Oza},
  \citenamefont {Ristroph},\ and\ \citenamefont {Shelley}}]{oza2019lattices}%
  \BibitemOpen
  \bibfield  {author} {\bibinfo {author} {\bibfnamefont {A.~U.}\ \bibnamefont
  {Oza}}, \bibinfo {author} {\bibfnamefont {L.}~\bibnamefont {Ristroph}},\ and\
  \bibinfo {author} {\bibfnamefont {M.~J.}\ \bibnamefont {Shelley}},\
  }\href@noop {} {\bibfield  {journal} {\bibinfo  {journal} {Phys. Rev. X}\
  }\textbf {\bibinfo {volume} {9}},\ \bibinfo {pages} {041024} (\bibinfo {year}
  {2019})}\BibitemShut {NoStop}%
\bibitem [{\citenamefont {Scholz}\ \emph {et~al.}(2018)\citenamefont {Scholz},
  \citenamefont {Jahanshahi}, \citenamefont {Ldov},\ and\ \citenamefont
  {L\"{o}wen}}]{Scholz2018}%
  \BibitemOpen
  \bibfield  {author} {\bibinfo {author} {\bibfnamefont {C.}~\bibnamefont
  {Scholz}}, \bibinfo {author} {\bibfnamefont {S.}~\bibnamefont {Jahanshahi}},
  \bibinfo {author} {\bibfnamefont {A.}~\bibnamefont {Ldov}},\ and\ \bibinfo
  {author} {\bibfnamefont {H.}~\bibnamefont {L\"{o}wen}},\ }\href@noop {}
  {\bibfield  {journal} {\bibinfo  {journal} {Nat. Commun.}\ }\textbf {\bibinfo
  {volume} {9}},\ \bibinfo {pages} {5156} (\bibinfo {year} {2018})}\BibitemShut
  {NoStop}%
\bibitem [{\citenamefont {Mandal}\ \emph {et~al.}(2019)\citenamefont {Mandal},
  \citenamefont {Liebchen},\ and\ \citenamefont {L\"owen}}]{Mandal2019}%
  \BibitemOpen
  \bibfield  {author} {\bibinfo {author} {\bibfnamefont {S.}~\bibnamefont
  {Mandal}}, \bibinfo {author} {\bibfnamefont {B.}~\bibnamefont {Liebchen}},\
  and\ \bibinfo {author} {\bibfnamefont {H.}~\bibnamefont {L\"owen}},\ }\href
  {https://doi.org/10.1103/PhysRevLett.123.228001} {\bibfield  {journal}
  {\bibinfo  {journal} {Phys. Rev. Lett.}\ }\textbf {\bibinfo {volume} {123}},\
  \bibinfo {pages} {228001} (\bibinfo {year} {2019})}\BibitemShut {NoStop}%
\bibitem [{\citenamefont {Banerjee}\ \emph {et~al.}(2017)\citenamefont
  {Banerjee}, \citenamefont {Souslov}, \citenamefont {Abanov},\ and\
  \citenamefont {Vitelli}}]{Banerjee2017}%
  \BibitemOpen
  \bibfield  {author} {\bibinfo {author} {\bibfnamefont {D.}~\bibnamefont
  {Banerjee}}, \bibinfo {author} {\bibfnamefont {A.}~\bibnamefont {Souslov}},
  \bibinfo {author} {\bibfnamefont {A.~G.}\ \bibnamefont {Abanov}},\ and\
  \bibinfo {author} {\bibfnamefont {V.}~\bibnamefont {Vitelli}},\ }\href@noop
  {} {\bibfield  {journal} {\bibinfo  {journal} {Nat. Commun.}\ }\textbf
  {\bibinfo {volume} {8}} (\bibinfo {year} {2017})}\BibitemShut {NoStop}%
\bibitem [{\citenamefont {Klotsa}(2019)}]{klotsa2019above}%
  \BibitemOpen
  \bibfield  {author} {\bibinfo {author} {\bibfnamefont {D.}~\bibnamefont
  {Klotsa}},\ }\href@noop {} {\bibfield  {journal} {\bibinfo  {journal} {Soft
  Matter}\ }\textbf {\bibinfo {volume} {15}},\ \bibinfo {pages} {8946}
  (\bibinfo {year} {2019})}\BibitemShut {NoStop}%
\bibitem [{\citenamefont {Bechinger}\ \emph {et~al.}(2016)\citenamefont
  {Bechinger}, \citenamefont {{Di Leonardo}}, \citenamefont {L{\"{o}}wen},
  \citenamefont {Reichhardt}, \citenamefont {Volpe},\ and\ \citenamefont
  {Volpe}}]{bechinger2016active}%
  \BibitemOpen
  \bibfield  {author} {\bibinfo {author} {\bibfnamefont {C.}~\bibnamefont
  {Bechinger}}, \bibinfo {author} {\bibfnamefont {R.}~\bibnamefont {{Di
  Leonardo}}}, \bibinfo {author} {\bibfnamefont {H.}~\bibnamefont
  {L{\"{o}}wen}}, \bibinfo {author} {\bibfnamefont {C.}~\bibnamefont
  {Reichhardt}}, \bibinfo {author} {\bibfnamefont {G.}~\bibnamefont {Volpe}},\
  and\ \bibinfo {author} {\bibfnamefont {G.}~\bibnamefont {Volpe}},\
  }\href@noop {} {\bibfield  {journal} {\bibinfo  {journal} {Rev. Mod. Phys.}\
  }\textbf {\bibinfo {volume} {88}},\ \bibinfo {pages} {45006} (\bibinfo {year}
  {2016})}\BibitemShut {NoStop}%
\bibitem [{\citenamefont {Dorbolo}\ \emph {et~al.}(2005)\citenamefont
  {Dorbolo}, \citenamefont {Volfson}, \citenamefont {Tsimring},\ and\
  \citenamefont {Kudrolli}}]{dorbolo2005dynamics}%
  \BibitemOpen
  \bibfield  {author} {\bibinfo {author} {\bibfnamefont {S.}~\bibnamefont
  {Dorbolo}}, \bibinfo {author} {\bibfnamefont {D.}~\bibnamefont {Volfson}},
  \bibinfo {author} {\bibfnamefont {L.}~\bibnamefont {Tsimring}},\ and\
  \bibinfo {author} {\bibfnamefont {A.}~\bibnamefont {Kudrolli}},\ }\href@noop
  {} {\bibfield  {journal} {\bibinfo  {journal} {Phys. Rev. Lett.}\ }\textbf
  {\bibinfo {volume} {95}},\ \bibinfo {pages} {044101} (\bibinfo {year}
  {2005})}\BibitemShut {NoStop}%
\bibitem [{\citenamefont {Couder}\ \emph {et~al.}(2005)\citenamefont {Couder},
  \citenamefont {Protiere}, \citenamefont {Fort},\ and\ \citenamefont
  {Boudaoud}}]{couder2005walking}%
  \BibitemOpen
  \bibfield  {author} {\bibinfo {author} {\bibfnamefont {Y.}~\bibnamefont
  {Couder}}, \bibinfo {author} {\bibfnamefont {S.}~\bibnamefont {Protiere}},
  \bibinfo {author} {\bibfnamefont {E.}~\bibnamefont {Fort}},\ and\ \bibinfo
  {author} {\bibfnamefont {A.}~\bibnamefont {Boudaoud}},\ }\href@noop {}
  {\bibfield  {journal} {\bibinfo  {journal} {Nature}\ }\textbf {\bibinfo
  {volume} {437}},\ \bibinfo {pages} {208} (\bibinfo {year}
  {2005})}\BibitemShut {NoStop}%
\bibitem [{\citenamefont {Narayan}\ \emph {et~al.}(2007)\citenamefont
  {Narayan}, \citenamefont {Ramaswamy},\ and\ \citenamefont
  {Menon}}]{narayan2007long}%
  \BibitemOpen
  \bibfield  {author} {\bibinfo {author} {\bibfnamefont {V.}~\bibnamefont
  {Narayan}}, \bibinfo {author} {\bibfnamefont {S.}~\bibnamefont {Ramaswamy}},\
  and\ \bibinfo {author} {\bibfnamefont {N.}~\bibnamefont {Menon}},\
  }\href@noop {} {\bibfield  {journal} {\bibinfo  {journal} {Science}\ }\textbf
  {\bibinfo {volume} {317}},\ \bibinfo {pages} {105} (\bibinfo {year}
  {2007})}\BibitemShut {NoStop}%
\bibitem [{\citenamefont {Aranson}\ \emph {et~al.}(2007)\citenamefont
  {Aranson}, \citenamefont {Volfson},\ and\ \citenamefont
  {Tsimring}}]{aranson2007swirling}%
  \BibitemOpen
  \bibfield  {author} {\bibinfo {author} {\bibfnamefont {I.~S.}\ \bibnamefont
  {Aranson}}, \bibinfo {author} {\bibfnamefont {D.}~\bibnamefont {Volfson}},\
  and\ \bibinfo {author} {\bibfnamefont {L.~S.}\ \bibnamefont {Tsimring}},\
  }\href@noop {} {\bibfield  {journal} {\bibinfo  {journal} {Phys. Rev. E}\
  }\textbf {\bibinfo {volume} {75}},\ \bibinfo {pages} {051301} (\bibinfo
  {year} {2007})}\BibitemShut {NoStop}%
\bibitem [{\citenamefont {Kudrolli}\ \emph {et~al.}(2008)\citenamefont
  {Kudrolli}, \citenamefont {Lumay}, \citenamefont {Volfson},\ and\
  \citenamefont {Tsimring}}]{kudrolli2008swarming}%
  \BibitemOpen
  \bibfield  {author} {\bibinfo {author} {\bibfnamefont {A.}~\bibnamefont
  {Kudrolli}}, \bibinfo {author} {\bibfnamefont {G.}~\bibnamefont {Lumay}},
  \bibinfo {author} {\bibfnamefont {D.}~\bibnamefont {Volfson}},\ and\ \bibinfo
  {author} {\bibfnamefont {L.~S.}\ \bibnamefont {Tsimring}},\ }\href@noop {}
  {\bibfield  {journal} {\bibinfo  {journal} {Phys. Rev. Lett.}\ }\textbf
  {\bibinfo {volume} {100}},\ \bibinfo {pages} {058001} (\bibinfo {year}
  {2008})}\BibitemShut {NoStop}%
\bibitem [{\citenamefont {Deseigne}\ \emph {et~al.}(2010)\citenamefont
  {Deseigne}, \citenamefont {Dauchot},\ and\ \citenamefont
  {Chat{\'e}}}]{deseigne2010collective}%
  \BibitemOpen
  \bibfield  {author} {\bibinfo {author} {\bibfnamefont {J.}~\bibnamefont
  {Deseigne}}, \bibinfo {author} {\bibfnamefont {O.}~\bibnamefont {Dauchot}},\
  and\ \bibinfo {author} {\bibfnamefont {H.}~\bibnamefont {Chat{\'e}}},\
  }\href@noop {} {\bibfield  {journal} {\bibinfo  {journal} {Phys. Rev. Lett.}\
  }\textbf {\bibinfo {volume} {105}},\ \bibinfo {pages} {098001} (\bibinfo
  {year} {2010})}\BibitemShut {NoStop}%
\bibitem [{\citenamefont {Workamp}\ \emph {et~al.}(2018)\citenamefont
  {Workamp}, \citenamefont {Ramirez}, \citenamefont {Daniels},\ and\
  \citenamefont {Dijksman}}]{Workamp_Disks}%
  \BibitemOpen
  \bibfield  {author} {\bibinfo {author} {\bibfnamefont {M.}~\bibnamefont
  {Workamp}}, \bibinfo {author} {\bibfnamefont {G.}~\bibnamefont {Ramirez}},
  \bibinfo {author} {\bibfnamefont {K.~E.}\ \bibnamefont {Daniels}},\ and\
  \bibinfo {author} {\bibfnamefont {J.~A.}\ \bibnamefont {Dijksman}},\
  }\href@noop {} {\bibfield  {journal} {\bibinfo  {journal} {Soft Matter}\
  }\textbf {\bibinfo {volume} {14}} (\bibinfo {year} {2018})}\BibitemShut
  {NoStop}%
\bibitem [{\citenamefont {S{\'{a}}enz}\ \emph {et~al.}(2018)\citenamefont
  {S{\'{a}}enz}, \citenamefont {Pucci}, \citenamefont {Goujon}, \citenamefont
  {Cristea-Platon}, \citenamefont {Dunkel},\ and\ \citenamefont
  {Bush}}]{saenz2018spin}%
  \BibitemOpen
  \bibfield  {author} {\bibinfo {author} {\bibfnamefont {P.~J.}\ \bibnamefont
  {S{\'{a}}enz}}, \bibinfo {author} {\bibfnamefont {G.}~\bibnamefont {Pucci}},
  \bibinfo {author} {\bibfnamefont {A.}~\bibnamefont {Goujon}}, \bibinfo
  {author} {\bibfnamefont {T.}~\bibnamefont {Cristea-Platon}}, \bibinfo
  {author} {\bibfnamefont {J.}~\bibnamefont {Dunkel}},\ and\ \bibinfo {author}
  {\bibfnamefont {J.~W.~M.}\ \bibnamefont {Bush}},\ }\href@noop {} {\bibfield
  {journal} {\bibinfo  {journal} {Phys. Rev. Fluids}\ }\textbf {\bibinfo
  {volume} {3}},\ \bibinfo {pages} {100508} (\bibinfo {year}
  {2018})}\BibitemShut {NoStop}%
\bibitem [{\citenamefont {Giomi}\ \emph {et~al.}(2013)\citenamefont {Giomi},
  \citenamefont {Hawley-Weld},\ and\ \citenamefont
  {Mahadevan}}]{giomi2013swarming}%
  \BibitemOpen
  \bibfield  {author} {\bibinfo {author} {\bibfnamefont {L.}~\bibnamefont
  {Giomi}}, \bibinfo {author} {\bibfnamefont {N.}~\bibnamefont {Hawley-Weld}},\
  and\ \bibinfo {author} {\bibfnamefont {L.}~\bibnamefont {Mahadevan}},\
  }\href@noop {} {\bibfield  {journal} {\bibinfo  {journal} {Proc. R. Soc. A}\
  }\textbf {\bibinfo {volume} {469}},\ \bibinfo {pages} {20120637} (\bibinfo
  {year} {2013})}\BibitemShut {NoStop}%
\bibitem [{\citenamefont {Deblais}\ \emph {et~al.}(2018)\citenamefont
  {Deblais}, \citenamefont {Barois}, \citenamefont {Guerin}, \citenamefont
  {Delville}, \citenamefont {Vaudaine}, \citenamefont {Lintuvuori},
  \citenamefont {Boudet}, \citenamefont {Baret},\ and\ \citenamefont
  {Kellay}}]{deblais2018boundaries}%
  \BibitemOpen
  \bibfield  {author} {\bibinfo {author} {\bibfnamefont {A.}~\bibnamefont
  {Deblais}}, \bibinfo {author} {\bibfnamefont {T.}~\bibnamefont {Barois}},
  \bibinfo {author} {\bibfnamefont {T.}~\bibnamefont {Guerin}}, \bibinfo
  {author} {\bibfnamefont {P.-H.}\ \bibnamefont {Delville}}, \bibinfo {author}
  {\bibfnamefont {R.}~\bibnamefont {Vaudaine}}, \bibinfo {author}
  {\bibfnamefont {J.~S.}\ \bibnamefont {Lintuvuori}}, \bibinfo {author}
  {\bibfnamefont {J.-F.}\ \bibnamefont {Boudet}}, \bibinfo {author}
  {\bibfnamefont {J.-C.}\ \bibnamefont {Baret}},\ and\ \bibinfo {author}
  {\bibfnamefont {H.}~\bibnamefont {Kellay}},\ }\href@noop {} {\bibfield
  {journal} {\bibinfo  {journal} {Phys. Rev. Lett.}\ }\textbf {\bibinfo
  {volume} {120}},\ \bibinfo {pages} {188002} (\bibinfo {year}
  {2018})}\BibitemShut {NoStop}%
\bibitem [{\citenamefont {Yamada}\ \emph {et~al.}(2003)\citenamefont {Yamada},
  \citenamefont {Hondou},\ and\ \citenamefont {Sano}}]{yamada2003coherent}%
  \BibitemOpen
  \bibfield  {author} {\bibinfo {author} {\bibfnamefont {D.}~\bibnamefont
  {Yamada}}, \bibinfo {author} {\bibfnamefont {T.}~\bibnamefont {Hondou}},\
  and\ \bibinfo {author} {\bibfnamefont {M.}~\bibnamefont {Sano}},\ }\href@noop
  {} {\bibfield  {journal} {\bibinfo  {journal} {Phys. Rev. E}\ }\textbf
  {\bibinfo {volume} {67}},\ \bibinfo {pages} {040301(R)} (\bibinfo {year}
  {2003})}\BibitemShut {NoStop}%
\bibitem [{\citenamefont {Koumakis}\ \emph {et~al.}(2016)\citenamefont
  {Koumakis}, \citenamefont {Gnoli}, \citenamefont {Maggi}, \citenamefont
  {Puglisi},\ and\ \citenamefont {Di~Leonardo}}]{koumakis2016mechanism}%
  \BibitemOpen
  \bibfield  {author} {\bibinfo {author} {\bibfnamefont {N.}~\bibnamefont
  {Koumakis}}, \bibinfo {author} {\bibfnamefont {A.}~\bibnamefont {Gnoli}},
  \bibinfo {author} {\bibfnamefont {C.}~\bibnamefont {Maggi}}, \bibinfo
  {author} {\bibfnamefont {A.}~\bibnamefont {Puglisi}},\ and\ \bibinfo {author}
  {\bibfnamefont {R.}~\bibnamefont {Di~Leonardo}},\ }\href@noop {} {\bibfield
  {journal} {\bibinfo  {journal} {New J. Phys.}\ }\textbf {\bibinfo {volume}
  {18}},\ \bibinfo {pages} {113046} (\bibinfo {year} {2016})}\BibitemShut
  {NoStop}%
\bibitem [{\citenamefont {Chat{\'{e}}}(2019)}]{chate2019dry}%
  \BibitemOpen
  \bibfield  {author} {\bibinfo {author} {\bibfnamefont {H.}~\bibnamefont
  {Chat{\'{e}}}},\ }\href@noop {} {\bibfield  {journal} {\bibinfo  {journal}
  {Annu. Rev. Condens. Matter Phys.}\ }\textbf {\bibinfo {volume} {11}}
  (\bibinfo {year} {2019})}\BibitemShut {NoStop}%
\bibitem [{\citenamefont {Bush}\ and\ \citenamefont
  {Hu}(2006)}]{bush2006walking}%
  \BibitemOpen
  \bibfield  {author} {\bibinfo {author} {\bibfnamefont {J.~W.~M.}\
  \bibnamefont {Bush}}\ and\ \bibinfo {author} {\bibfnamefont {D.~L.}\
  \bibnamefont {Hu}},\ }\href@noop {} {\bibfield  {journal} {\bibinfo
  {journal} {Annu. Rev. Fluid Mech.}\ }\textbf {\bibinfo {volume} {38}},\
  \bibinfo {pages} {339} (\bibinfo {year} {2006})}\BibitemShut {NoStop}%
\bibitem [{\citenamefont {Nagayama}\ \emph {et~al.}(2004)\citenamefont
  {Nagayama}, \citenamefont {Nakata}, \citenamefont {Doi},\ and\ \citenamefont
  {Hayashima}}]{nagayama2004theoretical}%
  \BibitemOpen
  \bibfield  {author} {\bibinfo {author} {\bibfnamefont {M.}~\bibnamefont
  {Nagayama}}, \bibinfo {author} {\bibfnamefont {S.}~\bibnamefont {Nakata}},
  \bibinfo {author} {\bibfnamefont {Y.}~\bibnamefont {Doi}},\ and\ \bibinfo
  {author} {\bibfnamefont {Y.}~\bibnamefont {Hayashima}},\ }\href@noop {}
  {\bibfield  {journal} {\bibinfo  {journal} {Phys. D Nonlinear Phenom.}\
  }\textbf {\bibinfo {volume} {194}},\ \bibinfo {pages} {151} (\bibinfo {year}
  {2004})}\BibitemShut {NoStop}%
\bibitem [{\citenamefont {Snezhko}\ \emph {et~al.}(2009)\citenamefont
  {Snezhko}, \citenamefont {Belkin}, \citenamefont {Aranson},\ and\
  \citenamefont {Kwok}}]{snezhko2009self}%
  \BibitemOpen
  \bibfield  {author} {\bibinfo {author} {\bibfnamefont {A.}~\bibnamefont
  {Snezhko}}, \bibinfo {author} {\bibfnamefont {M.}~\bibnamefont {Belkin}},
  \bibinfo {author} {\bibfnamefont {I.~S.}\ \bibnamefont {Aranson}},\ and\
  \bibinfo {author} {\bibfnamefont {W.-K.}\ \bibnamefont {Kwok}},\ }\href@noop
  {} {\bibfield  {journal} {\bibinfo  {journal} {Phys. Rev. Lett.}\ }\textbf
  {\bibinfo {volume} {102}},\ \bibinfo {pages} {118103} (\bibinfo {year}
  {2009})}\BibitemShut {NoStop}%
\bibitem [{\citenamefont {Chung}\ \emph {et~al.}(2009)\citenamefont {Chung},
  \citenamefont {Ryu},\ and\ \citenamefont {Cho}}]{chung2009electrowetting}%
  \BibitemOpen
  \bibfield  {author} {\bibinfo {author} {\bibfnamefont {S.~K.}\ \bibnamefont
  {Chung}}, \bibinfo {author} {\bibfnamefont {K.}~\bibnamefont {Ryu}},\ and\
  \bibinfo {author} {\bibfnamefont {S.~K.}\ \bibnamefont {Cho}},\ }\href@noop
  {} {\bibfield  {journal} {\bibinfo  {journal} {Appl. Phys. Lett.}\ }\textbf
  {\bibinfo {volume} {95}},\ \bibinfo {pages} {014107} (\bibinfo {year}
  {2009})}\BibitemShut {NoStop}%
\bibitem [{\citenamefont {Grosjean}\ \emph {et~al.}(2015)\citenamefont
  {Grosjean}, \citenamefont {Lagubeau}, \citenamefont {Darras}, \citenamefont
  {Hubert}, \citenamefont {Lumay},\ and\ \citenamefont
  {Vandewalle}}]{grosjean2015remote}%
  \BibitemOpen
  \bibfield  {author} {\bibinfo {author} {\bibfnamefont {G.}~\bibnamefont
  {Grosjean}}, \bibinfo {author} {\bibfnamefont {G.}~\bibnamefont {Lagubeau}},
  \bibinfo {author} {\bibfnamefont {A.}~\bibnamefont {Darras}}, \bibinfo
  {author} {\bibfnamefont {M.}~\bibnamefont {Hubert}}, \bibinfo {author}
  {\bibfnamefont {G.}~\bibnamefont {Lumay}},\ and\ \bibinfo {author}
  {\bibfnamefont {N.}~\bibnamefont {Vandewalle}},\ }\href@noop {} {\bibfield
  {journal} {\bibinfo  {journal} {Sci. Rep.}\ }\textbf {\bibinfo {volume}
  {5}},\ \bibinfo {pages} {16035} (\bibinfo {year} {2015})}\BibitemShut
  {NoStop}%
\bibitem [{\citenamefont {Yang}\ \emph {et~al.}(2019)\citenamefont {Yang},
  \citenamefont {Davoodianidalik}, \citenamefont {Xia}, \citenamefont
  {Punzmann}, \citenamefont {Shats},\ and\ \citenamefont
  {Francois}}]{yang2019passive}%
  \BibitemOpen
  \bibfield  {author} {\bibinfo {author} {\bibfnamefont {J.}~\bibnamefont
  {Yang}}, \bibinfo {author} {\bibfnamefont {M.}~\bibnamefont
  {Davoodianidalik}}, \bibinfo {author} {\bibfnamefont {H.}~\bibnamefont
  {Xia}}, \bibinfo {author} {\bibfnamefont {H.}~\bibnamefont {Punzmann}},
  \bibinfo {author} {\bibfnamefont {M.}~\bibnamefont {Shats}},\ and\ \bibinfo
  {author} {\bibfnamefont {N.}~\bibnamefont {Francois}},\ }\href@noop {}
  {\bibfield  {journal} {\bibinfo  {journal} {Phys. Rev. Fluids}\ }\textbf
  {\bibinfo {volume} {4}},\ \bibinfo {pages} {104608} (\bibinfo {year}
  {2019})}\BibitemShut {NoStop}%
\bibitem [{\citenamefont {Pucci}\ \emph {et~al.}(2011)\citenamefont {Pucci},
  \citenamefont {Fort}, \citenamefont {Ben~Amar},\ and\ \citenamefont
  {Couder}}]{pucci2011mutual}%
  \BibitemOpen
  \bibfield  {author} {\bibinfo {author} {\bibfnamefont {G.}~\bibnamefont
  {Pucci}}, \bibinfo {author} {\bibfnamefont {E.}~\bibnamefont {Fort}},
  \bibinfo {author} {\bibfnamefont {M.}~\bibnamefont {Ben~Amar}},\ and\
  \bibinfo {author} {\bibfnamefont {Y.}~\bibnamefont {Couder}},\ }\href@noop {}
  {\bibfield  {journal} {\bibinfo  {journal} {Phys. Rev. Lett.}\ }\textbf
  {\bibinfo {volume} {106}},\ \bibinfo {pages} {024503} (\bibinfo {year}
  {2011})}\BibitemShut {NoStop}%
\bibitem [{\citenamefont {Pucci}(2015)}]{pucci2015faraday}%
  \BibitemOpen
  \bibfield  {author} {\bibinfo {author} {\bibfnamefont {G.}~\bibnamefont
  {Pucci}},\ }\href@noop {} {\bibfield  {journal} {\bibinfo  {journal} {Int. J.
  Non Linear Mech.}\ }\textbf {\bibinfo {volume} {75}},\ \bibinfo {pages} {107}
  (\bibinfo {year} {2015})}\BibitemShut {NoStop}%
\bibitem [{\citenamefont {Ebata}\ and\ \citenamefont
  {Sano}(2015)}]{ebata2015swimming}%
  \BibitemOpen
  \bibfield  {author} {\bibinfo {author} {\bibfnamefont {H.}~\bibnamefont
  {Ebata}}\ and\ \bibinfo {author} {\bibfnamefont {M.}~\bibnamefont {Sano}},\
  }\href@noop {} {\bibfield  {journal} {\bibinfo  {journal} {Sci. Rep.}\
  }\textbf {\bibinfo {volume} {5}},\ \bibinfo {pages} {8546} (\bibinfo {year}
  {2015})}\BibitemShut {NoStop}%
\bibitem [{\citenamefont {Chen}\ \emph {et~al.}(2018)\citenamefont {Chen},
  \citenamefont {Doshi}, \citenamefont {Goldberg}, \citenamefont {Wang},\ and\
  \citenamefont {Wood}}]{chen2018controllable}%
  \BibitemOpen
  \bibfield  {author} {\bibinfo {author} {\bibfnamefont {Y.}~\bibnamefont
  {Chen}}, \bibinfo {author} {\bibfnamefont {N.}~\bibnamefont {Doshi}},
  \bibinfo {author} {\bibfnamefont {B.}~\bibnamefont {Goldberg}}, \bibinfo
  {author} {\bibfnamefont {H.}~\bibnamefont {Wang}},\ and\ \bibinfo {author}
  {\bibfnamefont {R.~J.}\ \bibnamefont {Wood}},\ }\href@noop {} {\bibfield
  {journal} {\bibinfo  {journal} {Nat. Commun.}\ }\textbf {\bibinfo {volume}
  {9}},\ \bibinfo {pages} {2495} (\bibinfo {year} {2018})}\BibitemShut
  {NoStop}%
\bibitem [{\citenamefont {Bush}\ and\ \citenamefont
  {Oza}(2020)}]{bush2020pilot}%
  \BibitemOpen
  \bibfield  {author} {\bibinfo {author} {\bibfnamefont {J.~W.~M.}\
  \bibnamefont {Bush}}\ and\ \bibinfo {author} {\bibfnamefont {A.~U.}\
  \bibnamefont {Oza}},\ }\href@noop {} {\bibfield  {journal} {\bibinfo
  {journal} {Rep. Prog. Phys.}\ }\textbf {\bibinfo {volume} {84}},\ \bibinfo
  {pages} {017001} (\bibinfo {year} {2020})}\BibitemShut {NoStop}%
\bibitem [{\citenamefont {Steinmann}\ \emph {et~al.}(2018)\citenamefont
  {Steinmann}, \citenamefont {Arutkin}, \citenamefont {Cochard}, \citenamefont
  {Rapha{\"e}l}, \citenamefont {Casas},\ and\ \citenamefont
  {Benzaquen}}]{steinmann2018unsteady}%
  \BibitemOpen
  \bibfield  {author} {\bibinfo {author} {\bibfnamefont {T.}~\bibnamefont
  {Steinmann}}, \bibinfo {author} {\bibfnamefont {M.}~\bibnamefont {Arutkin}},
  \bibinfo {author} {\bibfnamefont {P.}~\bibnamefont {Cochard}}, \bibinfo
  {author} {\bibfnamefont {E.}~\bibnamefont {Rapha{\"e}l}}, \bibinfo {author}
  {\bibfnamefont {J.}~\bibnamefont {Casas}},\ and\ \bibinfo {author}
  {\bibfnamefont {M.}~\bibnamefont {Benzaquen}},\ }\href@noop {} {\bibfield
  {journal} {\bibinfo  {journal} {J. Fluid Mech.}\ }\textbf {\bibinfo {volume}
  {848}},\ \bibinfo {pages} {370} (\bibinfo {year} {2018})}\BibitemShut
  {NoStop}%
\bibitem [{\citenamefont {Roh}\ and\ \citenamefont
  {Gharib}(2019)}]{roh2019honeybees}%
  \BibitemOpen
  \bibfield  {author} {\bibinfo {author} {\bibfnamefont {C.}~\bibnamefont
  {Roh}}\ and\ \bibinfo {author} {\bibfnamefont {M.}~\bibnamefont {Gharib}},\
  }\href@noop {} {\bibfield  {journal} {\bibinfo  {journal} {Proc. Natl. Acad.
  Sci.}\ }\textbf {\bibinfo {volume} {116}},\ \bibinfo {pages} {24446}
  (\bibinfo {year} {2019})}\BibitemShut {NoStop}%
\bibitem [{\citenamefont {Wilcox}(1972)}]{wilcox1972communication}%
  \BibitemOpen
  \bibfield  {author} {\bibinfo {author} {\bibfnamefont {R.~S.}\ \bibnamefont
  {Wilcox}},\ }\href@noop {} {\bibfield  {journal} {\bibinfo  {journal} {J.
  Comp. Physiol.}\ }\textbf {\bibinfo {volume} {80}},\ \bibinfo {pages} {255}
  (\bibinfo {year} {1972})}\BibitemShut {NoStop}%
\bibitem [{\citenamefont {Bleckmann}\ \emph {et~al.}(1994)\citenamefont
  {Bleckmann}, \citenamefont {Borchardt}, \citenamefont {Horn},\ and\
  \citenamefont {G{\"o}rner}}]{bleckmann1994stimulus}%
  \BibitemOpen
  \bibfield  {author} {\bibinfo {author} {\bibfnamefont {H.}~\bibnamefont
  {Bleckmann}}, \bibinfo {author} {\bibfnamefont {M.}~\bibnamefont
  {Borchardt}}, \bibinfo {author} {\bibfnamefont {P.}~\bibnamefont {Horn}},\
  and\ \bibinfo {author} {\bibfnamefont {P.}~\bibnamefont {G{\"o}rner}},\
  }\href@noop {} {\bibfield  {journal} {\bibinfo  {journal} {J. Comp. Phys.}\
  }\textbf {\bibinfo {volume} {174}},\ \bibinfo {pages} {305} (\bibinfo {year}
  {1994})}\BibitemShut {NoStop}%
\bibitem [{\citenamefont {Wright}\ and\ \citenamefont
  {Saylor}(2003)}]{wright2003patterning}%
  \BibitemOpen
  \bibfield  {author} {\bibinfo {author} {\bibfnamefont {P.}~\bibnamefont
  {Wright}}\ and\ \bibinfo {author} {\bibfnamefont {J.}~\bibnamefont
  {Saylor}},\ }\href@noop {} {\bibfield  {journal} {\bibinfo  {journal} {Rev.
  Sci. Instrum.}\ }\textbf {\bibinfo {volume} {74}},\ \bibinfo {pages} {4063}
  (\bibinfo {year} {2003})}\BibitemShut {NoStop}%
\bibitem [{\citenamefont {Falkovich}\ \emph {et~al.}(2005)\citenamefont
  {Falkovich}, \citenamefont {Weinberg}, \citenamefont {Denissenko},\ and\
  \citenamefont {Lukaschuk}}]{falkovich2005floater}%
  \BibitemOpen
  \bibfield  {author} {\bibinfo {author} {\bibfnamefont {G.}~\bibnamefont
  {Falkovich}}, \bibinfo {author} {\bibfnamefont {A.}~\bibnamefont {Weinberg}},
  \bibinfo {author} {\bibfnamefont {P.}~\bibnamefont {Denissenko}},\ and\
  \bibinfo {author} {\bibfnamefont {S.}~\bibnamefont {Lukaschuk}},\ }\href@noop
  {} {\bibfield  {journal} {\bibinfo  {journal} {Nature}\ }\textbf {\bibinfo
  {volume} {435}},\ \bibinfo {pages} {1045} (\bibinfo {year}
  {2005})}\BibitemShut {NoStop}%
\bibitem [{\citenamefont {Punzmann}\ \emph {et~al.}(2014)\citenamefont
  {Punzmann}, \citenamefont {Francois}, \citenamefont {Xia}, \citenamefont
  {Falkovich},\ and\ \citenamefont {Shats}}]{punzmann2014generation}%
  \BibitemOpen
  \bibfield  {author} {\bibinfo {author} {\bibfnamefont {H.}~\bibnamefont
  {Punzmann}}, \bibinfo {author} {\bibfnamefont {N.}~\bibnamefont {Francois}},
  \bibinfo {author} {\bibfnamefont {H.}~\bibnamefont {Xia}}, \bibinfo {author}
  {\bibfnamefont {G.}~\bibnamefont {Falkovich}},\ and\ \bibinfo {author}
  {\bibfnamefont {M.}~\bibnamefont {Shats}},\ }\href@noop {} {\bibfield
  {journal} {\bibinfo  {journal} {Nat. Phys.}\ }\textbf {\bibinfo {volume}
  {10}},\ \bibinfo {pages} {658} (\bibinfo {year} {2014})}\BibitemShut
  {NoStop}%
\bibitem [{\citenamefont {Rhee}\ \emph {et~al.}(2022)\citenamefont {Rhee},
  \citenamefont {Hunt}, \citenamefont {Thomson},\ and\ \citenamefont
  {Harris}}]{surferbot}%
  \BibitemOpen
  \bibfield  {author} {\bibinfo {author} {\bibfnamefont {E.}~\bibnamefont
  {Rhee}}, \bibinfo {author} {\bibfnamefont {R.}~\bibnamefont {Hunt}}, \bibinfo
  {author} {\bibfnamefont {S.~J.}\ \bibnamefont {Thomson}},\ and\ \bibinfo
  {author} {\bibfnamefont {D.~M.}\ \bibnamefont {Harris}},\ }\href@noop {}
  {\bibfield  {journal} {\bibinfo  {journal} {Bioinspiration \& Biomimetics}\
  }\textbf {\bibinfo {volume} {17}},\ \bibinfo {pages} {055001} (\bibinfo
  {year} {2022})}\BibitemShut {NoStop}%
\bibitem [{\citenamefont {Faraday}(1831)}]{faraday1831forms}%
  \BibitemOpen
  \bibfield  {author} {\bibinfo {author} {\bibfnamefont {M.}~\bibnamefont
  {Faraday}},\ }\href@noop {} {\bibfield  {journal} {\bibinfo  {journal}
  {Philos. Trans. R. Soc. London}\ }\textbf {\bibinfo {volume} {121}},\
  \bibinfo {pages} {319} (\bibinfo {year} {1831})}\BibitemShut {NoStop}%
\bibitem [{\citenamefont {Sanl{\i}}\ \emph {et~al.}(2014)\citenamefont
  {Sanl{\i}}, \citenamefont {Lohse},\ and\ \citenamefont {van~der
  Meer}}]{sanli2014antinode}%
  \BibitemOpen
  \bibfield  {author} {\bibinfo {author} {\bibfnamefont {C.}~\bibnamefont
  {Sanl{\i}}}, \bibinfo {author} {\bibfnamefont {D.}~\bibnamefont {Lohse}},\
  and\ \bibinfo {author} {\bibfnamefont {D.}~\bibnamefont {van~der Meer}},\
  }\href@noop {} {\bibfield  {journal} {\bibinfo  {journal} {Phys. Rev. E}\
  }\textbf {\bibinfo {volume} {89}},\ \bibinfo {pages} {053011} (\bibinfo
  {year} {2014})}\BibitemShut {NoStop}%
\bibitem [{sup()}]{supmat}%
  \BibitemOpen
  \href@noop {} {\emph {\bibinfo {title} {See Supplemental Material at [URL
  will be inserted by publisher] for supplementary figures, videos, and
  additional details on experimental methods and the theoretical
  model.}}}\BibitemShut {Stop}%
\bibitem [{\citenamefont {Longuet-Higgins}\ and\ \citenamefont
  {Stewart}(1964)}]{longuet1964radiation}%
  \BibitemOpen
  \bibfield  {author} {\bibinfo {author} {\bibfnamefont {M.~S.}\ \bibnamefont
  {Longuet-Higgins}}\ and\ \bibinfo {author} {\bibfnamefont {R.~W.}\
  \bibnamefont {Stewart}},\ }\href@noop {} {\bibfield  {journal} {\bibinfo
  {journal} {Deep-{S}ea Research}\ }\textbf {\bibinfo {volume} {11}},\ \bibinfo
  {pages} {529} (\bibinfo {year} {1964})}\BibitemShut {NoStop}%
\bibitem [{\citenamefont {Dombrowski}\ \emph {et~al.}(2022)\citenamefont
  {Dombrowski}, \citenamefont {Nguyen},\ and\ \citenamefont
  {Klotsa}}]{dombrowski2022pairwise}%
  \BibitemOpen
  \bibfield  {author} {\bibinfo {author} {\bibfnamefont {T.}~\bibnamefont
  {Dombrowski}}, \bibinfo {author} {\bibfnamefont {H.}~\bibnamefont {Nguyen}},\
  and\ \bibinfo {author} {\bibfnamefont {D.}~\bibnamefont {Klotsa}},\
  }\href@noop {} {\bibfield  {journal} {\bibinfo  {journal} {Physical Review
  Fluids}\ }\textbf {\bibinfo {volume} {7}},\ \bibinfo {pages} {074401}
  (\bibinfo {year} {2022})}\BibitemShut {NoStop}%
\bibitem [{\citenamefont {Ho}\ \emph {et~al.}(2019)\citenamefont {Ho},
  \citenamefont {Pucci},\ and\ \citenamefont {Harris}}]{ho2019direct}%
  \BibitemOpen
  \bibfield  {author} {\bibinfo {author} {\bibfnamefont {I.}~\bibnamefont
  {Ho}}, \bibinfo {author} {\bibfnamefont {G.}~\bibnamefont {Pucci}},\ and\
  \bibinfo {author} {\bibfnamefont {D.~M.}\ \bibnamefont {Harris}},\
  }\href@noop {} {\bibfield  {journal} {\bibinfo  {journal} {Phys. Rev. Lett.}\
  }\textbf {\bibinfo {volume} {123}},\ \bibinfo {pages} {254502} (\bibinfo
  {year} {2019})}\BibitemShut {NoStop}%
\bibitem [{\citenamefont {Suematsu}\ \emph {et~al.}(2010)\citenamefont
  {Suematsu}, \citenamefont {Nakata}, \citenamefont {Awazu},\ and\
  \citenamefont {Nishimori}}]{suematsu2010collective}%
  \BibitemOpen
  \bibfield  {author} {\bibinfo {author} {\bibfnamefont {N.~J.}\ \bibnamefont
  {Suematsu}}, \bibinfo {author} {\bibfnamefont {S.}~\bibnamefont {Nakata}},
  \bibinfo {author} {\bibfnamefont {A.}~\bibnamefont {Awazu}},\ and\ \bibinfo
  {author} {\bibfnamefont {H.}~\bibnamefont {Nishimori}},\ }\href@noop {}
  {\bibfield  {journal} {\bibinfo  {journal} {Phys. Rev. E}\ }\textbf {\bibinfo
  {volume} {81}},\ \bibinfo {pages} {056210} (\bibinfo {year}
  {2010})}\BibitemShut {NoStop}%
\end{thebibliography}%

\end{document}